\documentstyle[12pt,aaspp4,epsfig]{article}

\lefthead{Weinberger et al.}
\righthead{NGC 1068 Speckle}

\def\perarcsec{arcsec${-1}$}
\def\perpixel{pixel${-1}$}

\def\K{{\it K}$-$band~}
\def\H{{\it H}$-$band~}

\def\h{{\it H~}}
\def\k{{\it K~}}

\begin{document}

\title{Diffraction-Limited Imaging and Photometry of NGC 1068}

\author{A. J. Weinberger\altaffilmark{1}, G. Neugebauer, and K. Matthews}
\affil{Palomar Observatory, California Institute of Technology,
Pasadena, CA 91125\\
Electronic mail: alycia@astro.ucla.edu,
gxn@caltech.edu, kym@caltech.edu}

\altaffilmark{1}{Current Address: Department of Physics and Astronomy,
UCLA, Box 156205, Los Angeles, CA 90095}
\begin{abstract}

The nearby Seyfert 2 Galaxy NGC 1068 was observed with speckle imaging
techniques in the near-infrared \H (1.6 $\mu$m) at the Hale 200-inch
Telescope at Palomar Observatory and \K (2.2 $\mu$m) at the 10 m Keck I
Telescope at the W. M. Keck Observatory.

  Images with diffraction limited or near-diffraction limited
resolutions of 0.$''$05 $-$ 0.$''$1 were obtained and used to search for
structure in the nuclear region.  Images of the nucleus of NGC 1068
reveal an extended region of emission which accounts for nearly 50\% of
the nuclear flux at $K-$band.  This region extends 10 pc on either side
of an unresolved point source nucleus which is at most, 0.$''$02 or 1.4
pc in size.  Both the point source and the newly imaged extended
emission are very red, with identical $H-K$ colors corresponding to a
color temperature of 800~K.  While the point source is of a size to be
consistent with grains in thermal equilibrium with the nuclear source,
the extended emission is not. It must consist either of nuclear emission
which has been reflected off an extended dusty disk or of small grains
raised to transiently high temperatures by reflected UV photons.

\end{abstract}


\newpage

\section{Introduction}

As a close, luminous active galactic nucleus (AGN) [14.4 Mpc
(\cite{Tully88}) so that 1 arcsec=72 pc), NGC 1068 has been studied at
nearly every available spatial resolution and wavelength for thirty
years.  While classified as a Seyfert 2 based on the presence of narrow
emission lines and the absence of broad ones, polarization studies have
detected broad wings on its narrow emission lines (\cite{Ant85}).  These
observations suggest that NGC 1068 harbors an obscured Seyfert 1 nucleus
whose broad lines are scattered into our line of sight.  Significant
modeling of the spectrum and spectral energy distribution has been done
by Pier \& Krolik (1993) and Granato et al. (1997) for the presence of a
dusty torus which conceals the nucleus, and these models can reproduce
the observed emission from the X-ray through the near-infrared.

At scales of a few hundred parsecs around the nucleus, HST narrow band
(\cite{Macch94}) and continuum (\cite{Ly91}) imaging show a non-uniform
conical narrow-line region.  These observations suggest that clumps of
gas have been ionized by a partially collimated nuclear source.
Mid-infrared measurements of this area have revealed the presence of
warm gas (\cite{Cam93}).  Previous near infrared one$-$dimensional speckle
measurements of the nucleus (\cite{McCarthy82}; \cite{Chelli87}) showed extended
emission on the 100 pc scale, and more recent two$-$dimensional speckle
work finds extended emission closer to the nucleus (\cite{Wittkowski98}).
Radio VLBI measurements (\cite{Green96}) of water maser emission
demonstrate the presence of a thick torus, and high-resolution radio
maps of the nucleus (\cite{Galli96b}) suggest that the obscuring material
takes the form of a warped disk.

The combination of these results have made the nucleus of NGC 1068 the
prototypical obscured Seyfert 1.  In addition, the nucleus resides in an
SB host galaxy with a 3 kpc bar (\cite{Sco88}; \cite{Th89}) and active star
formation in the inner 10 kpc (\cite{Teles88}).  Authors have speculated
on the relationship between this star formation and the activity of the
nucleus (\cite{Normsco88}).

Near-infrared measurements trace the distribution of hot dust and stars
in NGC 1068, and thus characterize the physical condition of the
material near the nucleus.  The ability to do speckle imaging with the
W. M. Keck Telescope allows a resolution of 0.$''$05, or 3.6 pc, at 2.2
$\mu$m which for the first time provides a direct comparison between
near infrared and visual (HST) measurements.  We use these speckle
measurements, as well as complimentary 1.6 $\mu$m speckle imaging from
the 200-inch Telescope and direct imaging at both wavelengths from the
Keck Telescope to investigate the physical conditions in the near
nuclear region of NGC 1068 on hitherto unavailable scales.

\section{Observations}

Speckle observations of NGC 1068 were made on four nights, 1994 October
18, 1994 December 19, and 1995 November 4$-$5, with the 200-inch Hale
Telescope.  A 64$\times$64 subsection of a 256$\times$256 Santa Barbara
Research Center InSb array was used in order to allow continuous readout
of speckle frames every 0.07 or 0.10 s.  Speckle frames were collected
in sets of $\sim$400 images on the AGN and on the two nearby unresolved
SAO catalog stars.  The sources were observed at both \H
($\lambda_0$=1.65 $\mu$m, $\Delta\lambda$=0.32) and \K ($\lambda_0$=2.2
$\mu$m, $\Delta\lambda$=0.4).  Additional observations were made on
three nights, 1995 December 18-20, at the W. M. Keck Observatory. Images
from the full 256$\times$256 InSb array of the facility's Near Infrared
Camera (NIRC; \cite{MS94}), were taken at a rate of one 0.118~s image
every 1.5~s in sets of 100 images on the AGN and on the two nearby
unresolved stars SAO 130046 and SAO 110692. A \K ($\lambda_0$=2.21
$\mu$m, $\Delta\lambda$=0.43) filter was used for all the observations.
The basic observing strategy was reported in Matthews et al. (1996).  To
reduce the noise contributed by phase discontinuities between the 36
segments of the Keck Telescope, the object and calibrators were observed
at 12 different pupil orientations.  A summary of observations is
provided in Table \ref{tablejournal}.

\begin{table}[h]
\small
\caption{Journal of Observations of NGC 1068}
\medskip
\begin{center}
\begin{tabular}{cllcccccc}
\hline
Date &Cal 1 &Cal 2 &integ.       &\#frames &\# K &\# H &pixel scale &seeing \\
     &(SAO) &(SAO) &time (s)     &per set  &sets &sets &(arcsec     &(arcsec) \\
     &      &      &            &          &     &     &$\rm pixel^{-1}$) &\\
\hline
\hline
\multicolumn{9}{c}{200-Inch Telescope}\\
     &      &      &            &         &     &     &    & \\
18 Oct 1994     & 130057  & 110709   &0.1    &400    &11     &10     &0.036  &0.7 \\
19 Dec 1994     & 130057  & 110709   &0.1    &400    &20     &10     &0.036  &0.7 \\
4 Nov 1995      & 130057  & 110692   &0.07   &480    &14     &20     &0.034  &0.7 \\
5 Nov 1995      & 130057  & 110692   &0.1    &480    &16     &14     &0.034  &0.9 \\
     &      &      &            &         &     &     &    & \\
\multicolumn{9}{c}{Keck Telescope}\\
     &      &      &            &         &     &     &    & \\
18 Dec 1995      & 130046  & 110692   &0.118  &100    &11     &$-$ &0.021 &0.5 \\
19 Dec 1995      & 130046  & 110692   &0.118  &100    &12     &$-$ &0.021 &0.5 \\
20 Dec 1995      & 130046  & 110692   &0.118  &100    &16     &$-$ &0.021 &0.4 \\
16 Jan 1998     &$-$     &$-$      &10.0   &1      &3      &4       &0.021 &0.4 \\
16 Jan 1998     &$-$     &$-$      &5.0    &1      &1 &$-$ &0.15   &0.4 \\
\hline
\end{tabular}
\end{center}
\label{tablejournal}
\end{table}

At both telescopes, reimaging optics were used to convert the standard
detector plate scales to scales appropriate for diffraction limited
imaging.  At the 200-inch Hale Telescope, detector pixel scales of
0.$''$034 and 0.$''$036 \perpixel were used and at the Keck Telescope, a
detector pixel scale of 0.$''$021 \perpixel was used.  The pixel scale
at the 200-inch Telescope was chosen to oversample the \K while still
allowing diffraction limited imaging in $H-$band.  The pixel scale at
the Keck Telescope was chosen to sample optimally the aperture in the
$K-$band.

Several long exposure images of the nuclear region were also taken at
the Keck Telescope under photometric conditions in both the \H and
$K-$band.  In order to avoid saturating the detector, the speckle plate
scale of 0.$''$021 \perpixel was used for these images.  The total
integration times were 30 seconds at \K and 40 seconds at $H-$band, and the
seeing for these images was 0.$''$45.  A single 5 second \K image of NGC
1068 was also obtained at the Keck Telescope with a pixel scale of
0.$''$15/pixel and seeing of 0.$''$45.  Although the center is
saturated, this image captured the distribution of galactic \K flux at
distances greater than 0.$''$75 from the nucleus.  HST infrared standard
stars of (\cite{Persson}) were observed both with and without the
reimaging optics.

\section{Data Analysis}

In preliminary processing, each image was sky subtracted and flat
fielded, and bad pixels were corrected by interpolation.  The full
NIRC frame of 256$\times$256 pixels was clipped around the centroid of
each speckle frame to 128$\times$128 pixels (2.$''$6 on a side).
Outside this smaller field, the signal-to-noise-ratio in each pixel is
less than one-fifth, so clipping cut out pixels which would add only
noise to the Fourier analysis. In the 200-inch data no clipping was
necessary since a smaller field was used in collecting the images.  In
the second stage of analysis, the object's Fourier amplitudes and
phases were recovered via classical speckle analysis (\cite{Labey70})
and bispectral analysis (\cite{Weig87}) respectively.  For the data from
the Keck Telescope, both processes were modified from the standard
procedure to incorporate the field rotation that occurred during the
observations (\cite{Matth96}).  The observations were made over a spread
of 103$^\circ$ in parallactic angle, although the change in
parallactic angle over a single stack of 100 frames was always less
than $\sim$2$^\circ$.  Linear interpolation was used to find the
rotated pixel values in each frame.  The bispectral analysis was
sufficiently computationally intensive as to require the use of the
Caltech Concurrent Supercomputing Facility's nCUBE2 and Intel Delta
computers.

In order to go from the Fourier components calculated above to a final
image, it is necessary to include a smoothing function, or effectively a
telescope transfer function; a Gaussian of FWHM equal to the $\lambda /
D$ where $\lambda$ is the wavelength of the observations and $D$ is the
diameter of the telescope, was multiplied by the Fourier amplitudes.
Then, the amplitudes and phases were combined directly in an inverse
transform to produce the final image.

\section{Results} \label{results}

\subsection{$K-$band Results from the Keck Telescope} \label{kfit}

\begin{figure}[ht]
\label{fig_Keckimage}
\figcaption[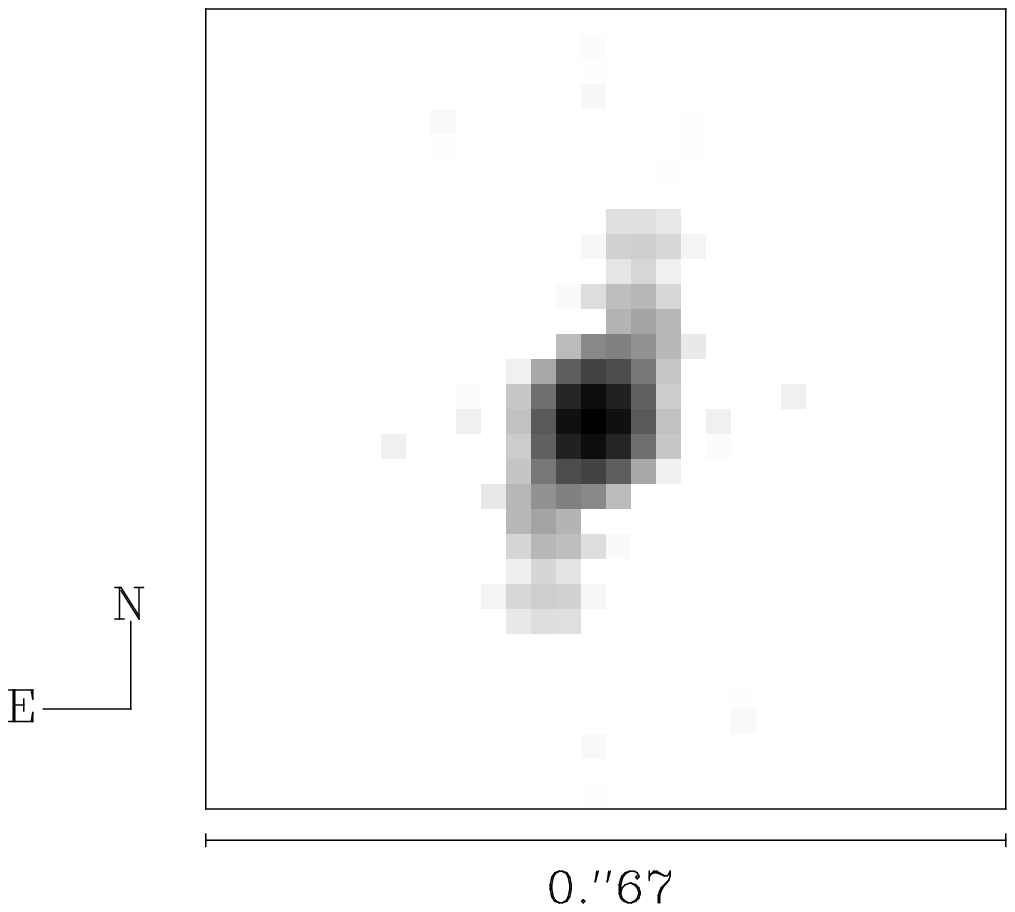]{Nucleus of NGC 1068 at 0.$''$05 (3.6 pc)
resolution.  Pixels are 0.$''$021 on a side, and the field of view has
been clipped to 0.$''$67 square.  The image is shown in a logarithmic
stretch and is comprised of two components -- a point source containing
49\% of the total nuclear flux and an extended component, lying at a
position angle of 159$^\circ$ and containing 51\% of the total nuclear
flux.}
\bigskip
\epsfig{file=fig1.ps}
\end{figure}

Figure \ref{fig_Keckimage} presents the 0.$''$05 (or 3.6 pc) resolution
\K image produced with speckle imaging from using all of the nearly 4000
frames obtained in 1995 December.  Each pixel is 0.$''$021 $\times$
0.$''$021 and the field of view has been clipped to 0.$''$67 on a side.
The nuclear emission is seen to be comprised of two components, an
unresolved point source and an extended region symmetric about the
nucleus with a major axis of $\sim$0.$''$3 (22 pc) and a minor axis of
$\sim$0.$''$18 (13 pc).  The calculated Fourier phases were consistent
with zero, so the extended component is symmetric about the nucleus.

\placefigure{fig_Keckimage}

For comparison with previous results and with models of the nucleus, we
estimated the fractions of the total nuclear flux density arising in the
point source and extended components.  These quantities, along with
the orientation and size of the extended emission, were found by fitting
a two component model to the two-dimensional average object visibility.
The fit was performed in the spatial frequency, i.e. Fourier, domain
rather than the image domain so that no tapering function had to be
applied to the high spatial frequencies.  This model was intended not to
reproduce the exact distribution of flux but to provide a robust
estimate of the magnitude of the contributions of the two components.

The extended emission was modeled as a smoothly falling exponential of
the form $I e^{-kd^n}$. The parameter $k$ measures the size of the
extended emission, and the parameter $d$ measures the shape and
orientation of the extended emission; $d$ was parameterized by an
ellipticity and angle.  The power $n$ determines how quickly the
visibility falls with spatial frequency and hence the overall shape of
the extended emission.  The power $n$ was assumed, by ad-hoc
phenomenological inspection, to be $3$ for the Keck Telescope \K
visibility.  Since this model has only the explicit purpose of measuring
the contributions from the two components, it is important for it to fit
the overall shape of the visibility but unimportant whether it
reproduced the details of the visibility at low spatial frequency.  In
particular, the lowest spatial frequencies were not used in the fit both
because they are the most corrupted by global changes in the seeing in
the time between when the object and calibrator were measured and
because the extended flux from the galaxy causes the visibility to drop
at the low frequencies.

\begin{figure}[ht]
\label{fig_1068_visib}
\figcaption[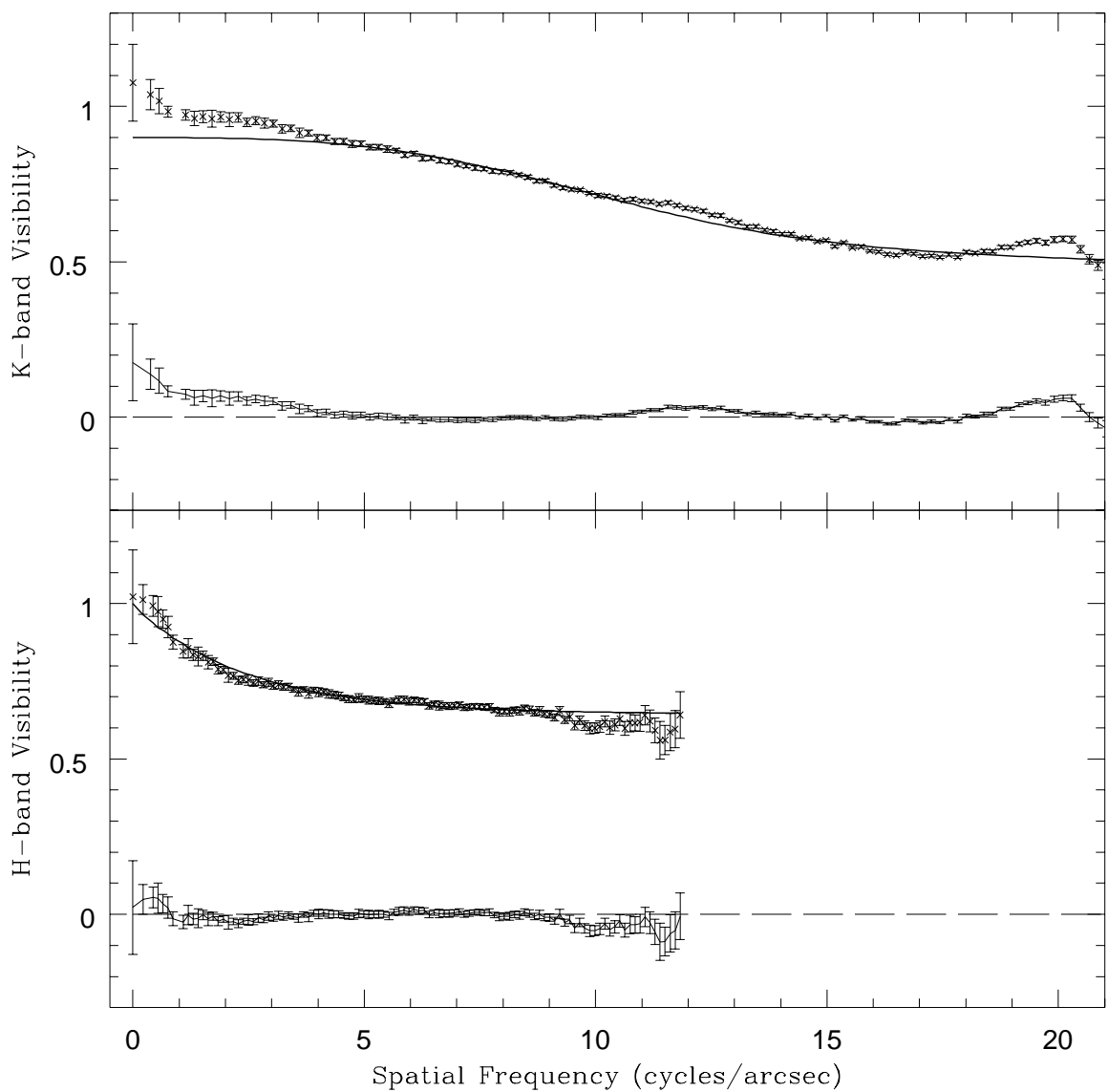]{Azimuthal averages of visibilities in
the $K-$band (top) and $H-$band (bottom). Measured visibilities are
represented by crosses in the top portion of each panel, with the line
indicating the two-component fit described in \S~\ref{kfit} or
\S~\ref{hfit}. Shown in the bottom portion of each panel are the
residuals (i.e. the data after the model has been subtracted). The line
through the residuals is at zero level.  Although the extended emission
is not azimuthally symmetric (see Figure 1), the radial averages provide
a convenient method of visual comparison between the data and the model
fit.}
\medskip
\epsfig{file=fig2.ps,height=6.5in,clip=}
\end{figure}

The point source was included as a constant visibility offset,
i.e. the same visibility at all spatial frequencies.  The fit was
performed as a $\chi^2$ minimization of the two-dimensional object
visibility, where each frequency was weighted by its statistical
uncertainty as calculated from the ensemble of power spectra, subject
to the constraints that the parameters be greater than or equal to
zero. All of the points from 0$-$20 cycles arcsec$^{-1}$ were included
in the fit. The salient results of this model are given in Table
\ref{tablefit}. Figure \ref{fig_1068_visib} shows the radial
(i.e. azimuthally averaged) profiles of the measured two-dimensional
visibilities and of the fit are shown.  The differences between the
data and model, i.e. the residuals, are also shown in the figure.

\placefigure{fig_1068_visib}

The result of the fitting demonstrates that 49\% of the \K flux density
in the speckle image is contained in the unresolved core, i.e. in a
diffraction limited beam, and 51\% is in the extended region.  The
uncertainty in this fit is taken to be the uncertainty in the
normalization of the power spectra, or 8\%.  \S~\ref{color}
describes how the total flux in the speckle image was determined.

\subsection{$H-$band Results from the Palomar 200-inch Telescope}
\label{hfit}

Of all the \H data taken at the 200-inch Telescope, the observations
from 19 December 1994 had the best signal to noise ratio at high spatial
frequencies, so they were used for the analysis which follows.  The
observations from other nights are consistent with that of 19 December,
but of lower quality.  The final \H image from the 200-inch Telescope,
with half the resolution of the \K image from the Keck Telescope
(0.$''$1 or 7.2 pc), similarly consists of both a point source and
extended emission.

The same fitting procedure described in the previous section was used to
fit the two-dimensional \H visibility, but the power $n$ was taken to be
$1$.  Since the value of $n$ primarily affects the shape of the
visibility at low spatial frequencies (where, as noted above, the
measurements are sensitive to seeing variation and the galactic
emission) we make no comparison between the \H and \K data on these
scales.

In the \H data from the 200-inch Telescope, 63\% of the flux density
from the speckle image is contained in the unresolved point source
(i.e. within a diffraction limited beam) and 37\% is in the extended
region.  The uncertainty in this fit is taken to be the uncertainty in
the normalization of the visibilities, or 8\%.  A discussion of how the
total flux in the \H image is determined is given in
\S~\ref{color}.

\begin{table}[ht]
\small
\caption{Results of Model Fits to 2-D Object Visibility}
\medskip
\begin{center}
\begin{tabular}{lccc}
\hline
measurement &eccentricity &angle &fraction unresolved \\
\hline
\hline
Keck \K (at 200-inch Resolution) &0.81	&158.2	&0.64 \\
200-Inch \H     		 &0.81	&159.4	&0.63 \\
Keck \K (at full Resolution)     &0.81	&158.6	&0.49 \\
\hline
\end{tabular}
\end{center}
\label{tablefit}
\end{table}

\subsection{\h$-$\k Color}

The data from the Keck Telescope, at higher spatial resolution than that
from the 200-inch Telescope, resolves more of the nuclear \K flux into
extended emission.  In order to compute the \h$-$\k colors of the
unresolved point source and the extended emission, however, the \K data
from the Keck Telescope must be smoothed to the 200-inch Telescope
resolution.  Instead of smoothing the reconstructed image, the object
visibility from the Keck Telescope data was fit out to a spatial
frequency of 11.3 cycles \perarcsec, i.e. the same resolution obtainable
at the 200-inch Telescope, using the procedure described above in
\S~\ref{kfit}.  In this case the \K shows the same distribution of flux
density as the $H-$band, i.e. 64\% in an unresolved core and 36\% in an
extended region.  Since both the \K and \H data show the same fraction
of their respective total fluxes in the point source, the H$-$K color of
the point source is the same as that of the extended emission.  The
uncertainty in this color is 11\%, the combination of the uncertainties
in each of the fits to the visibilities. The actual value of this color
is computed below in the discussion section.

\subsection{Upper Limit to the Size of the Point-like Nucleus} \label{ptsrc}

The speckle data can be used to place an upper limit on the size of
the nuclear point source.  If the core were actually extended, it would
have the effect of reducing the visibilities at high spatial frequencies,
but instead we find the visibilities flatten out at high spatial frequencies.
The fit to the Keck Telescope data, shown in a radial profile plot in
Figure \ref{fig_1068_visib}, leaves residuals of less than 10\% at
frequencies above 19 cycles~\perarcsec.  So, the presence of another
undetected component is therefore constrained by the uncertainty in the
residuals at the highest frequencies.  Without making an a priori assumption
as to the shape of the extension, should it be present, the upper limit
to its size can be taken as the highest spatial frequency at which high
S/N information was obtained in the data, in this case 19.7
cycles~\perarcsec or 0.$''$051.  A more stringent upper limit to the point
source size can be placed under the assumption that the true nucleus has
a Gaussian shape.  The width of the largest Gaussian which could be
hidden in the \K visibility data (i.e.  which will not differ from the
data by more than 3$\sigma$ at the highest spatial frequency) is
0.$''$02 or 1.4 pc.

\subsection{Photometry} \label{keckphot}

\begin{figure}[ht]
\label{fig_1068cog}
\figcaption[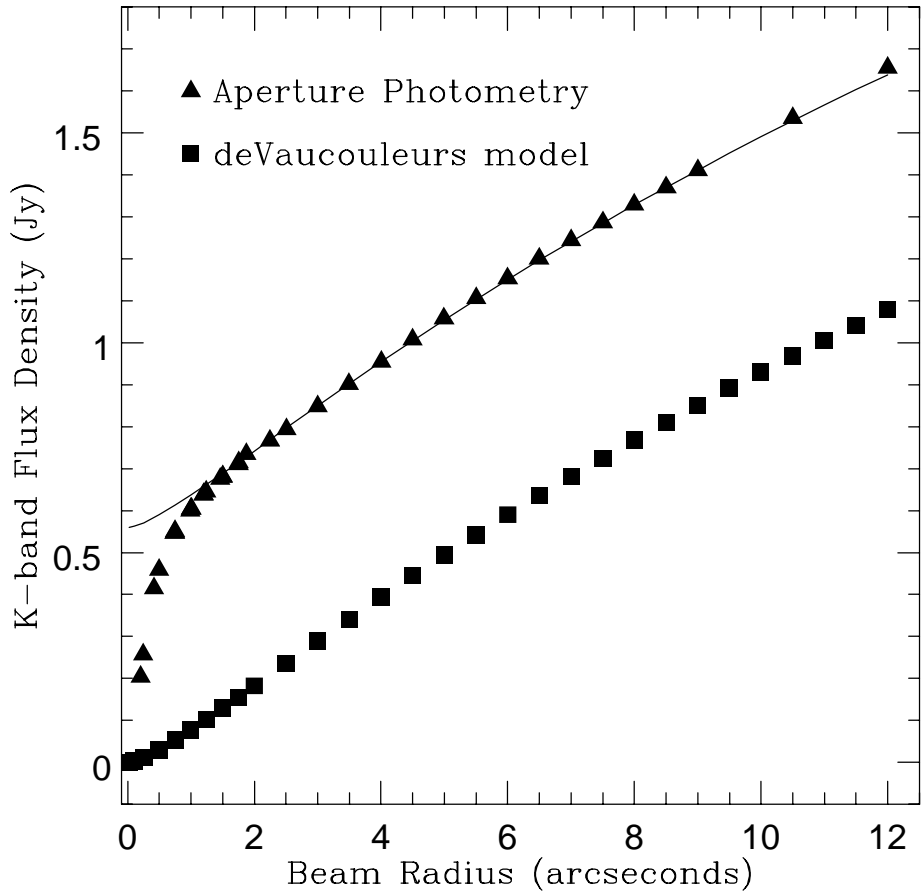]{Nuclear curve of growth, i.e. the \K
flux density in a series of successively larger beam sizes.  The points
on the top curve (triangles) are the data measured from the long
exposure images taken at the Keck Telescope, as described in
\S~\ref{keckphot}.  The seeing for these observations was 0.$''$45 FWHM,
so the drop off in the flux below radii of 1$''$ is due to seeing. The
line through the points is a sum of the deVaucouleurs model fit, done
using the measurements between radii 1.$''$8 and 27$''$, and a constant,
i.e. point source.  The lower curve (squares), are the deVaucouleurs
model fit at each radius for which photometry was
done.}
\bigskip
\epsfig{file=fig3.ps}
\end{figure}

Because of limitations deriving from the small field size, low S/N, and
image wander in speckle images, it is difficult to make photometric
measurements from speckle data.  Therefore, we measured the total flux
density in a beam the size of the speckle frames from 0.$''$45
resolution long exposure images made at the Keck Telescope in both the
\h and $K-$bands.  In a beam radius of 1.$''$25, the \K magnitude is 7.5
and the \H magnitude is 9.26, resulting in an {\it H}$-${\it K} color of
1.76~mag.  Aperture photometry from these images (in both magnitudes and
Janskys) at a variety of other beam sizes is reported in Table
\ref{table_photometry} and shown in Figure \ref{fig_1068cog}.

\placefigure{fig_1068cog}

The contribution of the galaxy to the photometry at small beam sizes was
estimated by fitting the galaxy surface brightness at radii between
1.$''$8 and 27$''$ with a deVaucouleurs function.  Extrapolating the fit
to a beam of radius 1.$''$25, approximately the same size as the speckle
field of view, implies that only 15\% percent of the flux arises in the
galaxy. Furthermore, the shape of the surface brightness profile within
1.$''$25 is consistent with being the sum of the deVaucouleurs profile
and a point source.

\begin{table}[ht]
\small
\caption{Aperture Photometry and Colors}
\medskip
\begin{center}
\begin{tabular}{cccccc}
\hline
beam diameter &K     &K     &H     &H     &H$-$K \\
($''$)        &(mag) &(mJy) &(mag) &(mJy) &(mag) \\
\hline
\hline
0.4    &8.75  &204    &11.06  &39   &2.31 \\
0.5    &8.50  &257    &10.78  &51   &2.28 \\
0.8    &7.98  &415    &10.13  &92   &2.14 \\
1.0    &7.87  &459    &9.95   &109  &2.08 \\
1.5    &7.68  &547    &9.63   &146  &1.95 \\
2.0    &7.58  &600    &9.42   &178  &1.84 \\
2.5    &7.50  &646    &9.26   &206  &1.76 \\
2.9    &7.45  &676    &9.14   &230  &1.69 \\
3.5    &7.39  &715    &9.00   &261  &1.61 \\
3.75   &7.36  &735    &8.94   &276  &1.58 \\
\hline
\end{tabular}
\end{center}
\label{table_photometry}

Uncertainties are 3\% in both the K and H magnitudes.  Within a diameter
of 1$''$, the uncertainties are systematically larger due to the effects
of the seeing disk.
\end{table}

\subsection{Comparison to Previous Measurements}

The size of the near-infrared core of NGC 1068 has been the subject of
investigation by many previous researchers.  One-dimensional speckle by
McCarthy et al. (1982) at the 3.8 m Mayall Telescope placed an upper limit
of 0.$''$2 on the size of the unresolved core, and found extended
emission that was 25\% of the total 2.2 $\mu$m flux in the beam.  As
shown in Figure \ref{fig_1068_visib}, the \K visibility measured in this
work is consistent with 1.0 out to 1 cycle~\perarcsec, but in the McCarthy
et al. data, the visibility decreased to 0.8 by 0.5 cycles~\perarcsec.
This is probably a consequence of their large, 5$''$ by 10$''$, beam,
which captured more light from the galactic stars, which have large
spatial extent, than our smaller 2.$''$6 square beam.

Similarly, 1-D speckle at 3.6 $\mu$m by Chelli et al. (1987) at the 3.6 m
ESO Telescope found an unresolved core, large scale (100 pc) emission,
and a third component of extended yet compact emission 0.$''$2 around
the nucleus.  The visibility obtained by Chelli et al. agrees very well
with what we report. Their data also suggested that, of the two position
angles they measured, the compact extended emission was larger along an
angle of 135$^\circ$ than 45$^\circ$ which is consistent with
the emission we measure shown in Figure \ref{fig_Keckimage}.

In more recent imaging with an aperture mask, Thatte et al. (1997) find
that 94\% of the \K flux in a 1$''$ diameter aperture comes from a
point source smaller than 0.$''$03.  They report the flux from this
source as 190 mJy.  Both of these measurements disagree with what is
reported above in \S~\ref{kfit} and Table~\ref{table_photometry},
respectively.

Recent two dimensional speckle imaging by Wittkowski et al. (1998) at the 6
m Special Astrophysical Observatory Telescope finds extended emission
which is 20\% of the total \K flux and in addition places a limit on the
core size of 0.$''$03.  These estimates came from assuming a uniform
disk model for the reconstructed emission, but they admit that an
alternative explanation for their data would be an unresolved central
object and extended emission. Fitting their data with this model, as
described in \S\S~\ref{kfit} and \ref{hfit} above, would increase
the fraction of the flux density attributable to the extended emission.
While of insufficient sensitivity to show the extended emission reported
in this work, their results are consistent with what is described here.

\section{Discussion}

Discussed in this section are three components of the emission from the
central 1.$''$25 radius of the nucleus: (1) the central point source,
(2) the newly imaged extended nuclear emission reported in
\S~\ref{results} (and accounting for approximately 50\% of the flux at
2$\mu$m previously attributed to the point source), and (3) the stars of
the underlying host galaxy.

\subsection{\h$-$\k Color of the Nucleus} \label{color}

While it is possible to tell from the speckle measurements alone that
the color of the point source and the color of the newly imaged
extended emission are the same, it is not possible to determine the
color itself.  This is because speckle, as an interferometric technique,
resolves out (i.e. is not sensitive to) smooth large-scale extended
emission which fills the field of view.  Thus, all of the flux measured
in a beam the size of the speckle frames, as reported in Table
\ref{table_photometry}, cannot be automatically attributed to the
features observed in the speckle image.  However, the color of the
emission in the speckle image can be deduced by subtracting the
contribution from large scale galactic emission from the total flux
measured in the speckle beam.  We assume that the only such contribution
comes from the distribution of stars in the host galaxy.

From the deVaucouleurs model fitting described in
\S~\ref{keckphot} the galactic stellar contribution to the 1.$''$25
radius speckle beam was determined to be 97 mJy at $K-$band, or 15\% of
the total \K emission.  Since no large (38$''$ square) image of NGC 1068
at \H such as the one taken at \K was available, it was assumed that the
$H-K$ color of the galactic stellar population was 0.3 mag in the
nuclear region, i.e. the same color as measured in aperture photometry
off the nucleus (\cite{Th89}).  Combining this color with the \K
measurement, it was calculated that 118 mJy, or 57\%, of the \H emission
in the 1.$''$25 radius beam is due to stars.  All flux in excess of the
galactic stellar contribution, was assumed to come from the nucleus plus
extended emission reported in \S~\ref{results}, and this flux then
has an $H-K$ color of 2.5 mag.  If there is another population of stars
in excess of the assumed galactic contribution, this estimate of the
$H-K$ color would be low.  If there is substantial reddening of the
stars in the nucleus compared with far from the nucleus, this estimate
would be high.  However, $H-K$~=~2.1~mag can safely be considered a
lower limit to the color based on the aperture photometry reported in
Table \ref{table_photometry}.  The statistical uncertainty in this
color is a combination of the uncertainties in the photometric
calibration, the aperture photometry, and the fit to the photometry, for
a total of 9\%.  Combined with the uncertainty in the visibility fitting
from \S\S~\ref{kfit} and \ref{hfit}, our best estimate of the color of the
extended emission is 2.5 $\pm$ 0.2 mag.

\subsection{Possible Mechanisms for Extended Emission} \label{extem}

It is of interest to consider the origin of the observed extended
nuclear emission.  There are several possibilities: it could be emission
from stars, from nuclear light reflected off dust or electrons, or
emission from hot dust, either in equilibrium or from single photon
heating.

The $H-K$ color of the extended emission, 2.5 mag, is significantly
redder than that of any stellar population.  Thus, if the emission is
from stars, it must be highly extincted.  A color excess of 2.2 mag,
obtained from assuming the stars have the same intrinsic color as the
galactic stars far from the nucleus, (i.e., $H-K$=0.3 mag) necessitates
$A_v$=34 mag.  This is similar to the extinction found in the center
parsec of our own Galaxy and produced in models of thermal emission from
dust in a thick torus in NGC 1068 (\cite{Efstat95}; \cite{Young95}).  It is,
however, much higher than the extinction of $A_v$=0$-$2 mags suggested
in the data of Thatte et al. (1997) for the central stellar cluster. It
is known, however, that the extinction is quite patchy near the nucleus
(\cite{Blietz94}).  The extended emission reported in this work is smooth
over a length of 20 pc, and the substantial reddening required for the
extended emission to be from stars seems unlikely to be similarly smooth
over this scale.

A bigger problem with this hypothesis comes from the luminosity of the
extended emission.  \K imaging spectroscopy of the nuclear region
(\cite{Thatte97}) has revealed the presence of a dense stellar cluster,
which, based on the equivalent width of the CO band-head, accounts for
7\% of the total nuclear luminosity.  However, 50\% of the nuclear flux
is resolved by speckle. The \K luminosity of the extended region, if it
is emitted isotropically, is 4.7$\times$10$^{8}$~L$_\odot$, a factor of
ten larger than the stellar luminosity given in Thatte et al.

Finally, in this scenario the fact that the point source and the
extended emission have the same color would be purely accidental unless
the point source is also composed of stars.  If the extended
near-infrared structure is comprised of stars, it is also unlikely to be
a continuation of a larger scale structure in the host galaxy. NGC 1068
does have a well known large scale ($\approx$ 1 kpc) stellar bar
(\cite{Sco88}; \cite{Th89}), but it is also oriented at a position angle of
approximately 45$^\circ$.

Light from the point source reflected off dust or electrons in the
narrow-line region would, on the other hand, provide a natural
explanation for the similar colors of the point source and extended
emission.  The nucleus is highly (4$-$5\%) polarized at 2.2$\mu$m in a
4$''$ beam suggesting that there is extensive scattering of the nuclear
radiation (\cite{Lebofsky78}). There are three possible sources of
scattering: the warm electron gas which scatters the broad line
emission, another population of electrons, or dust.  The warm electrons
modeled by Miller, Goodrich, \& Matthews (1991) are located at least 30 pc from the nucleus,
i.e.  outside the extended emission reported here. Therefore scattering
from these electrons is unlikely be the source of the extended
region.  Scattering from other electrons which are within a few parsecs
of the point source would also tend to reflect the broad line region, so
it is reasonable to conclude that there is no second population of
electrons beyond that found by Miller et al.

The albedo, and wavelength dependence thereof, of an ensemble of dust
grains varies widely with their size distribution
(e.g. \cite{Lehtinen96}).  For the small grains that are to be expected
in high UV-radiation field locations such as that around NGC 1068,
observations and theory (\cite{Draine84}) predict that the albedo at 2.2
$\mu$m is approximately 20\% lower than at 1.6 $\mu$m.  Therefore, if
the observed extended emission were reflected light from the point
source of NGC 1068, it would be significantly bluer than the point
source itself, whereas we observe the same color in the two sources.  On
the optimistic assumption that the albedo at 2.2 $\mu$m has a value of
0.8 and that the dust scatters isotropically, the central source would
have a true 2.2 $\mu$m luminosity which is 15 times greater than
observed.

Observations by Glass (1997) have shown that the nucleus of NGC 1068
became steadily brighter at \K over a twenty year span from 1976 to 1994
before leveling off.  He did not detect a concomitant rise in the \H
emission, but this is understandable given the galactic stellar
contamination of his 12$''$ beam.  If the nuclear emission comes from
dust on the inner edge of the torus which is heated to just below its
sublimation point, an increase in luminosity will push the inner edge of
the torus further away from the nucleus but not change the intrinsic
color of the emission.  However, the time constant for destroying grains
may be long enough, on the order of years (\cite{Voit92}) so that an
increase in luminosity produces temporarily higher temperatures and
therefore bluer colors.

The light crossing time of the extended emission reported in
\S~\ref{results} is approximately 10~yr, so if it is reflected point
source light, it should show a 10~yr lag in color compared to the point
source.  We do not know what the color of the point source was 10 years
ago, but for it to have been just red enough to offset the tendency of
dust reflection to make the emission bluer would be quite a conspiracy.

The final possibility is that the extended emission comes from hot dust.
The color temperature implied by an H$-$K color of 2.5 mags is 800 K.
Taking the central luminosity of the AGN as $1.5 \times
10^{11} L_{\sun}$, we can calculate that the dust grains which would be
heated to this temperature in equilibrium would lie at most 1 pc or
0.$''$01 (for silicate grains) from the point source.  The extended
emission reported in this paper is a factor of 10 larger.  It has been
suggested by other authors (\cite{Baldwin87}; \cite{Braatz93}; \cite{Bock98}) that
extended 10 $\mu$m emission on 200 pc scales may be caused by heating of
grains by the central source if the luminosity is beamed along the
direction of the radio jets rather than emitted isotropically.  The good
spatial correspondence between the mid-infrared and radio jet
(\cite{Cam93}; \cite{Bock98}) also lends credence to this idea.  The component of
the radio jet thought to lie at the infrared point source
(\cite{Galli96a}), S1, sits in a region of extended radio continuum
emission which lies at a position angle of 175$^\circ$.  A beaming
factor of 200 which would be sufficient to explain the extended
mid-infrared emission would also be sufficient to produce 800~K grains
at 10 pc from the point source.  By contrast, Efstathiou et al. (1995) derive a
beaming factor of $\gtrsim$~6 based on their fit to the near-infrared
spectrum of the nucleus, and this would be insufficient to heat grains
to 800~K, 10 pc from the point source.

It is possible that the extended near-infrared emission is from
hot dust which is heated not externally (i.e. by the central AGN), but
internally, for example, by an interaction with the
jet.  The jets are observed to drive the motion of the emission line gas
(\cite{Axon97}), so it is reasonable to assume that they are dumping
energy into the circumnuclear gas.  More complete models would have to
be made to examine this hypothesis.

A natural explanation for the high color temperature which is
reconcilable with an isotropically emitting central source would be
single photon heating of small grains (\cite{Sellgren84}).  The rate of
UV photons necessary to produce the total \K luminosity of the extended
nuclear emission can be calculated if one knows the mass of dust present
in the region.  If the hydrogen density is 10$^5$ cm$^{-3}$
(\cite{Tacconi94}), the dust to gas mass ratio is 10$^{-2}$, and the
grains radiate with unity efficiency in the infrared, the rate of
photons needed per grain is $\sim$10$^{-5}$~s$^{-1}$.  This rate is well
below that expected from the intrinsic UV/X-ray spectrum of NGC 1068
(\cite{Pier94}).  However, polycyclic aromatic hydrocarbons (PAHs) are
thought to be destroyed by intense X-ray/UV radiation fields
(\cite{Voit92}) and the 3.3 $\mu$m emission feature associated with PAHs
has not been unambiguously detected at the nucleus (\cite{BlandHaw97}).
However, if the dust along the edges of the extended torus, as in Figure
\ref{fig_1068model}, were illuminated by UV photons which were reflected
off of the high lying electron cloud and yet protected from the nucleus
by the bulk of the torus, the X-ray flux they intercept would be
substantially reduced.  Miller et al. (1991) predict that if the optical
depth to electron scattering is about 0.1 and there is a dusty disk of
dimension 10$^{20}$ cm surrounding the central region, then about 10\%
of the central UV luminosity would be back scattered onto the disk.  If
the disk is not uniform, as is likely considering the lumpy high
resolution radio maps, some regions would have high enough column
density to stop the grain destroying X-rays, yet see a reflected UV flux
sufficient to create transient grain heating.

\subsection{Comparison to Models and Line Emission}

\begin{figure}[ht]
\label{fig_1068model}
\figcaption[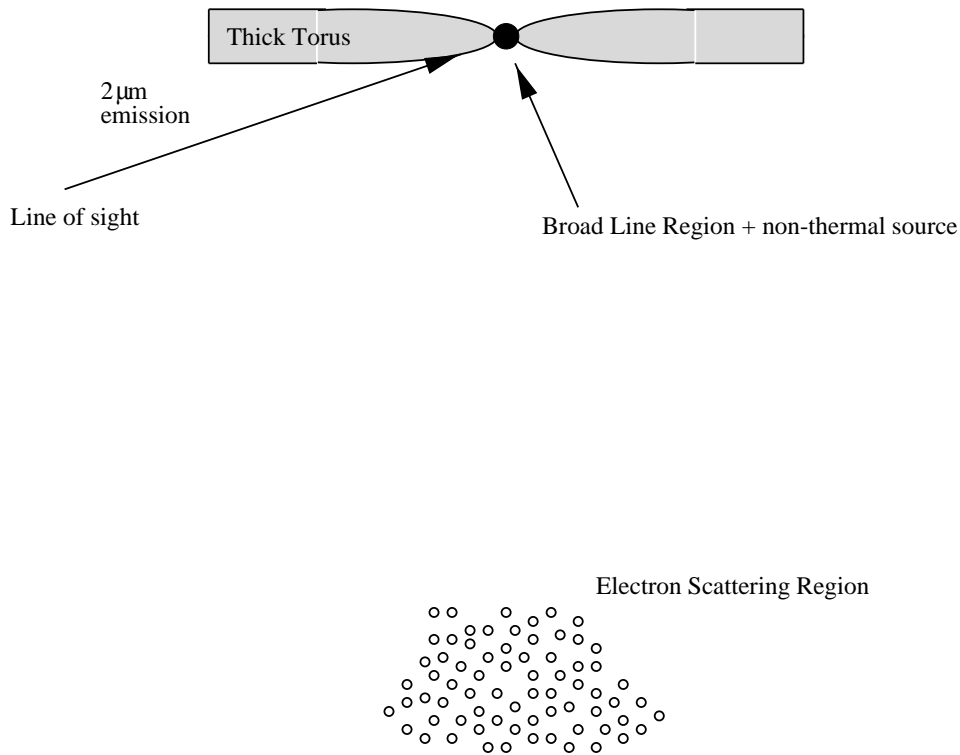] {Scale cartoon model of the nuclear
region of NGC 1068.  The black hole, accretion disk, and broad line
region are within the black circle of radius 0.2 pc which defines the
center.  The optically thick torus has an inner radius of $\sim$0.2 pc
set by the sublimation point of dust grains in the intense radiation
field of the point source. The distance along this torus at which grains
reach a temperature of 800 K if heated by a central source of luminosity
1.5$\times$10$^{11}$ L$_\odot$ is 1 pc. The electron clouds responsible
for reflecting the broad line region to the line of sight sit 30 pc
above and below the nucleus.}
\bigskip
\epsfig{file=fig4.ps,height=4in,clip=}
\end{figure}

In the model of infrared emission from NGC 1068
(e.g. \cite{pnk93}; \cite{Efstat95}; \cite{Gran97}) commonly found in the literature, the
central source is surrounded by an optically thick torus.  The torus is
heated by the ultraviolet and X-ray photons from the accretion disk plus
black hole system to which it is optically thick, but ionizing photons
escape along the axis of the torus.  The inner radius of the torus,
$\sim$0.2 pc, is set by the sublimation point of the dust, and its outer
radius, $\leq$ 40 pc (\cite{Gran97}), is set by models of its infrared
emission.  The line of sight to this torus is nearly edge on, passing
through 70-1000 magnitudes of visual extinction, depending on the model,
and therefore does not permit a direct view of the central source. The
1$-$2 $\mu$m emission observed is produced by the thermal radiation from
hot dust on the inner edge of the torus, which, because it is on the
edge, escapes through a region of moderate extinction.  The geometry of
the torus is constrained by the conical shape of the narrow-line region
to have an opening angle of approximately 45$^\circ$.  A cartoon of this
model is shown in Figure \ref{fig_1068model}.  The 2 pc upper limit
placed on the size of the point source in \S~\ref{ptsrc} is
consistent with this model, but the fact that the extended emission we
observe is much larger than the size of the inner edge of the torus
means that we must add to this picture.  The emission we observe at 10
pc from the nucleus could come from a larger scale dusty structure,
perhaps an extension of the torus, if the emission can be produced by
one of the mechanisms outlined in \S~\ref{extem}.

\placefigure{fig_1068model}

\begin{figure}[ht]
\label{fig_hst}
\figcaption[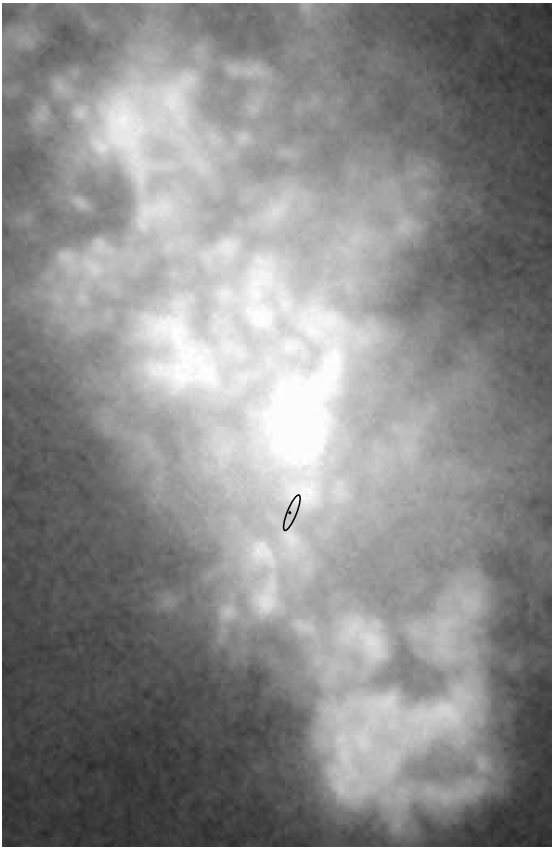]{Location and size of the extended \K
emission (ellipse), superposed with the HST $[$O III$]$ emission line
image of Macchetto et al. (1994). The infrared point source was located
at the center of polarization determined by Capetti et al. (1995).  The
\K flux does not arise from any of the bright clumps of material which
produce the narrow lines.}
\bigskip
\epsfig{file=fig5.ps,height=4in,clip=,angle=-90}
\end{figure}

Light scattering off of dust is observed in the narrow-line region much
further from the central source (\cite{MGM91}) than predicted by models of
the spectral energy distribution (\cite{Efstat95}).  The narrow-line
emission comes from clouds excited by the central source (\cite{Macch94}),
and it extends to hundreds of parsecs, at a position angle of
approximately 45$^\circ$.  The placement of the 2 $\mu$m point source
and extended region, shown superposed on the HST narrow-line image in
Figure \ref{fig_hst}, shows that the extended 2 $\mu$m emission lies
alongside the bright emission knots in the visual ionization cone.  Of
course, the registration of the infrared and optical images is not known
to exquisite precision. The best estimates from Thatte et al. (1997) have a
0.$''$1 uncertainty in the registration and this uncertainty nearly
encompasses the size of the extended near-infrared emission.

\placefigure{fig_hst}

\section{Conclusions}

Two components of the nuclear 2.2 and 1.6 $\mu$m emission of NGC 1068,
in addition to its galactic stellar population, have been detected with
speckle imaging on the Keck and 200-Inch Telescopes.  The observations
reveal an extended region of emission that accounts for nearly 50\% of
the nuclear flux at $K-$band.  This region extends 10 pc along its major
axis and 6 pc along its minor axis on either side of an unresolved point
source nucleus which is at most, 0.$''$02 or 1.4 pc in size.

Both the point source and the newly imaged extended emission are very
red, with identical $H-K$ colors corresponding to a color temperature of
800~K.  While the point source is of a size to be consistent with grains
in thermal equilibrium with the nuclear source, the extended emission is
not. The current data do not allow us to unambiguously determine the
origin of the extended emission, but it is most likely either scattered
nuclear radiation from an extended dusty disk or emission from thermally
fluctuating small grains heated by reflected nuclear UV photons.

\acknowledgements

We thank Andrea Ghez for her help with the observations and data
analysis, Tom Soifer for many helpful conversations on models of NGC
1068, and the telescope operators at Palomar and Keck Observatories for
their efforts during time consuming speckle observing.  Infrared
astronomy at Caltech is supported by the NSF.

\newpage

\end{document}